\documentclass[final,5p,times,twocolumn]{elsarticle}
\usepackage{graphicx}
\usepackage{slashed,color}
\usepackage{amsmath,amsfonts,bm}
\usepackage{amssymb}
\usepackage[T1]{fontenc}

\newcommand{\nn}{\nonumber \\ }

\journal{Physics Letters  B}

\begin{document}

%\begin{frontmatter}

%\preprint{JLAB-THY-17-2579,  October 24, 2017}

\author{A.~V.~Radyushkin}
\address{Physics Department, Old Dominion University, Norfolk,
             VA 23529, USA}
\address{Thomas Jefferson National Accelerator Facility,
              Newport News, VA 23606, USA
}

\title{Quark  Pseudodistributions  at Short Distances}

\begin{abstract}

%{Preprint JLAB-THY-17-2579}
%\bigskip

We perform a   one-loop study  of  the small-$z_3^2$
behavior of the Ioffe-time distribution (ITD)  ${\cal M} (\nu, z_3^2)$,  
 the basic function that may be converted 
 into parton 
pseudo-  and quasi-distributions. 
We calculate the   corrections on   the 
operator level, so that our  results may be also 
used for pseudo-distribution amplitudes and 
generalized parton pseudodistributions. 
We 
separate  two sources of the $z_3^2$-dependence 
at small   $z_3^2$. One  is related to the 
ultraviolet (UV) singularities generated by the gauge  link.
Our calculation  shows that, for a finite UV cut-off, the  UV-singular terms 
vanish   when  $z_3^2=0$. The  UV divergences are  absent  
in the  reduced ITD given by the  ratio ${\cal M} (\nu, z_3^2)/{\cal M} (0, z_3^2)$.
Still, it  has a non-trivial short-distance 
behavior due to   $\ln z_3^2  \Lambda^2$   terms generating 
  perturbative evolution of  parton densities.
 We give an  explicit expression, up to constant terms, 
for  the reduced ITD at one loop.  It may  be used 
in   extraction  of  PDFs from  lattice QCD  simulations. 
We  also use our results to get  new insights concerning   the
 structure of parton quasi-distributions at one-loop  level.

%\vspace{5mm} 

%Keywords: Parton distribution  functions; Transverse  momentum; Quasi-distributions
              
\end{abstract}

%\pacs{11.10.-z,12.38.-t,13.60.Fz}
\maketitle

%\end{frontmatter}

\section{
 Introduction}

The usual parton distribution functions (PDFs) $f(x)$ \cite{Feynman:1973xc}  
are often mentioned  now as ``light-cone PDFs'',
since they 
are  related to   matrix elements $M(z,p)$ of    the $\langle  p | \phi (0)  \phi(z)  | p \rangle $
type taken 
 on   the light cone $z^2=0$.
 The parton {\it pseudo-distributions}  ${\cal P} (x, -z^2)$  \cite{Radyushkin:2017cyf,Radyushkin:2017sfi} 
 generalize PDFs for a situation when $z$ is off the light cone,
 in particular, when $z$ is  spacelike $z^2 <  0 $.  To obtain them,
 one should treat $M(z,p) $ as a function  ${\cal M} (\nu, -z^2)$ 
 of the Ioffe time $(pz)\equiv  -\nu$ \cite{Ioffe:1969kf} 
 and $z^2$. Fourier-transforming  ${\cal M} (\nu, -z^2)$ with respect to $\nu$
 gives pseudo-PDFs ${\cal P} (x, -z^2)$.  
  
  As  suggested by X. Ji \cite{Ji:2013dva},  
  using  purely spacelike
separations $z=(0,0,0,z_3)$ (or, for brevity, $z=z_3$) allows  one to study 
matrix elements $M(z_3,p) $  (and hence, the 
Ioffe-time \mbox{distributions   \cite{Braun:1994jq} }   ${\cal M} (\nu, z_3^2)$)  on the lattice.
To extract PDFs from such studies, 
one needs  to  understand  the small-$z_3^2$ 
behavior of the pseudo-PDFs ${\cal P} (x, z_3^2)$.

In any renormalizable theory,
a perturbative calculation of 
the matrix element $M(z_3,p)$  
reveals   logarithmic $\ln z_3^2$   singularities
resulting in  the  perturbative evolution of PDFs. 
Further complications arise in 
  quantum chromodynamics (QCD), where   the parton fields
are connected by a gauge link $E(0,z;A)$. As emphasized by Polyakov
\cite{Polyakov:1980ca}, 
 perturbative corrections 
to gauge  links (and loops)  result in  specific ultraviolet  divergences
requiring an  additional renormalization
(see  also \cite{Dotsenko:1979wb,Brandt:1981kf,Aoyama:1981ev}).  %%%%%%%
Studies of the UV and short-distance properties 
of  the QCD bilocal operators $\bar \psi (0) \gamma^\alpha E(0,z;A) \psi (z)$
were performed
by Craigie and Dorn   %%%%%%
 in Ref. \cite{Craigie:1980qs}
 (see also  recent papers \cite{Ishikawa:2017faj,Ji:2017oey,Green:2017xeu}). %%%%%%%%

An important point is that the effects of perturbative 
gluon corrections may be formulated on   the operator level,
without  reference to a particular matrix element in which 
the operator is inserted. 
 This idea
   is most efficient when realized in the form of the nonlocal (or string) form of %%%%%%%
   light-cone operator product expansion  (OPE)  %%%%%%%%%%%%%
   \cite{Anikin:1978tj}  %%%%%%%%
%%%%  was implemented
that was developed %%%%%%%%%
 in much detail 
by Balitsky and Braun \cite{Balitsky:1987bk} who applied it 
to studies of the light-cone limit. For this reason, they skipped 
the discussion of the link-specific UV divergences that,
at a finite UV cut-off, %%%%%%%%
 disappear when $z^2=0$. 
 Also, they did  not present  the $z^2$-independent part of the one-loop contribution. %%%%%%%

In the present paper, we perform a 
complete
%%%%%
one-loop study of the QCD bilocal 
operators {\it off}  the light cone. 
In  the context of the quasi-distributions   approach \cite{Ji:2013dva},  %%%%%%
the one loop corrections have been recently calculated %%%%%%
for quasi-PDFs  \cite{Xiong:2013bka,Ji:2015jwa},   %%%%%%
for generalized parton quasi-distributions    %%%%%%
and for pion quasi-distribution amplitude  %%%%%%
\cite{Ji:2015qla}.  %%%%%%
All of them result from     independent calculations of  %%%%%%%%%
 Feynman diagrams in momentum space,   %%%%%%%%%
 specific for each  type of  quasi-distributions.   %%%%%%%%%
  The produced results have rather lengthy analytical forms, that   strongly %
  differ  for each  case. In particular,  the very question did not arise %
if, say, the result for the pion quasi-distribution amplitude  and quasi-GPDs  in Ref. \cite{Ji:2015qla}  %%%%%% %
are  consistent with  the result for the nonsinglet quasi-PDF in Ref.  \cite{Xiong:2013bka,Ji:2015jwa}. %%%%% %

In   contrast,  using the nonlocal OPE  requires just a single %%%%%%%
calculation of Feynman diagrams. The results obtained  in  the operator form  %%%
are rather compact, and may be converted into any parton distribution %%%%%%%
by a relevant Fourier transform.  %%%%%%%%%%
As examples, we present an expression  for the reduced ITD,   %%%%%%%
and we also show how a simple $\ln (z_3^2 m^2)$ evolution  logarithm %%%%%%%
proliferates into a rather complicated structure for quasi-PDF. %%%%%%%

Furthermore, working in  the coordinate representation allows one %%%%%%%
to clearly separate UV singular terms  from those governed by the %%%%%%%
evolution effects.  %%%%%%%%
We pay 
a special %%%%%%%
attention both to the 
UV-singular link-related contributions 
and to the UV-finite evolution-related  terms singular in the $z_3^2 \to 0$ limit.

Technically,  our calculations are similar to those
performed in our paper \cite{Radyushkin:2015gpa} within the formalism 
of virtuality distributions  \cite{Radyushkin:2014vla}.
In some aspects, our calculations  also resemble  those 
of Ref.  \cite{Balitsky:1987bk},   
where the calculations, however,  %%%%%%%%%
did not go beyond the leading evolution logarithm. %%%%%%%%%

 We start, in Section 2,  with a short  overview  of  the basic ideas of our approach formulated in Refs. 
\cite{Radyushkin:2016hsy,Radyushkin:2017cyf}. In Section 3, we  
outline the 
nonlocal OPE  %%%%%%%%
calculation   of the one-loop diagrams
in the operator form,  and identify the terms  %%%%%%%%%%
 responsible for 
the behavior of the ITD ${\cal M} (\nu, z_3^2)$ for small $z_3$. 
\mbox{In Section 4, }  we show that rather simple expressions for the one-loop 
ITD may be used for a  straightforward calculation 
of the one-loop quasi-PDFs, providing new insights concerning their 
structure.  Section 5  contains the  summary of the paper.

 %\setcounter{equation}{0}   {\it Parton distributions} 

%{\it DIS }

%\newpage

\section{PDFs}

 \subsection{Ioffe-time distributions}

The basic parton distribution functions (PDFs)  introduced by Feynman   \cite{Feynman:1973xc}
  are  extracted from  the 
matrix elements of bilocal operators, 
generically  written as   $\langle  p | \phi (0)  \phi(z)  | p \rangle $.
We use here scalar notations 
for the partonic  fields because 
 complications related to spin are not central to the very concept of PDFs. 
 
In many situations  (especially  in extraction of  parton distributions
from  the lattice),  it is very useful  to treat such a  matrix element 
 as a function 
${\cal M} (\nu, -z^2) $ of   two Lorentz invariants, the {\it Ioffe time}
\cite{Ioffe:1969kf}   \mbox{$(pz) \equiv -  \nu$}  
 and   the interval $-z^2$. Thus, we write 
  \begin{align}
  \langle p |   \phi(0) \phi (z)|p \rangle&  \equiv M(z,p) \nn & 
=   {\cal M} (-(pz), -z^2)  =  {\cal M} (\nu, -z^2)  
\,  .
 \
 \label{lorentz}
\end{align} 
The function 
    ${\cal M}(\nu , -z^2) $  is    the {\it Ioffe-time distribution}  \cite{Braun:1994jq}. 
    
 It can be shown \cite{Radyushkin:2016hsy,Radyushkin:1983wh}    that, for all contributing  Feynman diagrams,   
 the   Fourier transform  ${\cal P} (x, -z^2)$ of the ITD with respect to $\nu$ 
 has the $-1 \leq x \leq 1$ support, i.e., 
   \begin{align}
 {\cal M} (\nu, -z^2) 
&   = 
 \int_{-1}^1 dx 
 \, e^{i x \nu  } \,  {\cal P} (x, -z^2)  \   .
  \label{MPD}
\end{align}   
  Note that  in this covariant definition of $x$ we do not  assume that 
$z^2=0$ or $p^2=0$.

 \subsection{Light-cone    PDFs}
 
 The expressions for  various types of parton distributions,
 all in terms of the same ITD   ${\cal M} (-(pz), -z^2)$,  may be  obtained from 
special  choices of 
 $z$ and $p$. 
In particular, taking  a lightlike  $z$, e.g., that  
having  the light-front minus component $z_-$  only, we   
parameterize the matrix element  through  the twist-2 parton distribution $f(x)$ 
  \begin{align}
  {\cal M} (-p_+ z_- , 0)  =
   \int_{-1}^1 dx \, f(x) \, 
e^{-ixp_+ z_-} \,  \   . 
 \label{twist2par0}
\end{align}
It has  the usual interpretation of  probability that 
the parton carries  the fraction $x$ of the target momentum 
component $p_+$.  
The inverse relation is given by 
\begin{align} 
 f  (x)  =\frac{1}{2 \pi}  \int_{-\infty}^{\infty}  d\nu \, e^{-i x \nu}  \, {\cal M}(\nu , 0)    = {\cal P} (x, 0) 
  \  .
\label{fxMnu}
\end{align}  
Since $f(x) =   {\cal P} (x, 0) $,  we may  say that 
the function $ {\cal P} (x, -z^2) $ generalizes 
the concept of the usual
light-cone parton distribution onto intervals $z^2$ that are not light-like. 
The functions  $ {\cal P} (x, -z^2)  $  will be referred to as parton 
{\it pseudo-distributions} (or pseudo-PDFs) \cite{Radyushkin:2017cyf}.

 \subsection{Quasidistributions} 
Since one cannot have light-like separations on the lattice,
it was proposed  \cite{Ji:2013dva}  to consider 
equal-time  separations  
$z= (0,0,0,z_3)$ [or \mbox{$z=z_3$}, for brevity]. Then, 
 in  the   \mbox{$p=(E, 0_\perp, P)$}  frame,
 one  introduces quasi-PDF  $Q(y, P)$    through a parametrization  
   \begin{align}
  {\cal  M} (Pz_3, z_3^2)   =  & 
   \int_{-\infty}^{\infty}   {dy}   \, Q(y, P)\, 
 \,  e^{i y  P z_3 } \, 
 \  . 
 \label{MyPQ}
\end{align} 
The quasi-PDF  describes  the distribution of  parton's \mbox{$k_3=yP$}  momentum.
Using  the inverse relation 
  \begin{align}
   Q(y, P) =  & \frac{|P|}{2 \pi} 
   \int_{-\infty}^{\infty}   {dz_3}
 \,  e^{-i y  P z_3 } \, 
  {\cal  M} (Pz_3, z_3^2) 
 \  , 
 \label{QyP}
\end{align} 
we can express the quasi-PDF     $Q(y, P)$    in terms of the pseudo-PDF 
$ {\cal   P} (x, z_3^2) $ corresponding to the $z=z_3$ separation 
 \begin{align}
     Q(y,P) =\frac{|P|}{2 \pi}  \int_{-1}^1 dx \,   \int_{-\infty}^\infty dz_3\,
      e^{-i(y-x) Pz_3}\, 
 {\cal   P} (x, z_3^2)  \ . 
  \label{Py}
\end{align}

When ${\cal   P} (x, z_3^2)$ does not depend on $z_3$, i.e., when
${\cal   P} (x, z_3^2) =f(x)$, the quasidistribution $Q(y,P)$ does not depend on
$P$, and coincides with  the PDF $f(y)$.

It should be noted  that 
 the integration variable $z_3$  in Eq. (\ref{QyP}) enters  into both arguments of the 
ITD $  {\cal  M} (Pz_3, z_3^2) $.  In contrast, the pseudo-PDF 
$ {\cal P} (x, z_3^2) $ is obtained through integrating the ITD  
$ {\cal  M} (\nu, z_3^2)$  just with respect to its first argument, 
  \begin{align}
 {\cal P} (x, z_3^2)   
&   = \frac{1}{2 \pi}  
 \int_{-\infty}^\infty  d\nu
 \, e^{-i x \nu } \,  {\cal  M} (\nu, z_3^2)
  \   .
  \label{PMDx}
\end{align}

 \subsection{Transverse momentum dependent
distributions}   

If one treats 
the target momentum $p$ as   longitudinal,
\mbox{$p= (E, {\bf 0}_\perp, P)$,}   but 
chooses  $z$ that has a transverse 
 $z_\perp= \{z_1,z_2\}$ component, one can  
  introduce  the transverse momentum dependent distributions (TMDs). 
Namely, 
 taking $z_+=0$,    one defines  the  
{\it TMD }   ${\cal F} (x, k_\perp^2)$ by 
     \begin{align}
 {\cal M} (\nu,  z_\perp^2) 
&   = 
 \int_{-1}^1 dx \,\, e^{i x  \nu  }  
    \int_{-\infty}^\infty  d^2{\bf k}_\perp   e^{i ({\bf k}_\perp {\bf z}_\perp)} 
      {\cal F} (x, k_\perp^2)
 \   .
  \label{MTMD}
\end{align}   
 One  may also write 
    \begin{align}
 {\cal P} (x,  z_\perp^2) 
&   = 
    \int  d^2{\bf k}_\perp   e^{i ({\bf k}_\perp {\bf z}_\perp)} 
      {\cal F} (x, {k}_\perp^2)
 \   .
  \label{MTMD0}
\end{align}  
This means that 
 the pseudo-PDF   $ {\cal P} (x,  z_\perp^2) $ coincides
 in this case 
with the  {\it impact parameter distribution}, a   concept 
that is well known  
from TMD studies.

While $z=z_3$ corresponds to  a purely ``longitudinal'' 
separation, the Lorentz invariance  
requires that ${\cal P} (x,- z^2) $  is the same function 
of $-z^2$ no matter what kind of a space-like $z$ we deal with.
In other words, 
 the dependence of  ${\cal P} (x, z_3^2) $  
on its second argument reflects the same physics 
that leads to TMDs.

 \subsection{QCD case}  
 
 The formulas given above 
may be used in case of  nonsinglet parton densities   in QCD.
The relevant matrix elements  are 
    \begin{align}
 { M}^\alpha  (z,p) \equiv \langle  p |  \bar \psi (0) \,
 \gamma^\alpha \,  { \hat E} (0,z; A) \psi (z) | p \rangle \  , 
\end{align}
 where  $
{ \hat E}(0,z; A)$ is  the  standard  $0\to z$ straight-line gauge link 
 in the quark (fundamental) 
 representation
 \begin{align}
{ \hat E}(0,z; A) \equiv P \exp{ \left [ ig \,  z_\nu\, \int_0^1dt \,  \hat  A^\nu (t z) 
 \right ] }    \  . 
 \label{straightE}
\end{align}
 These matrix elements   have  $p^\alpha$ and $z^\alpha$ parts
    \begin{align}
{ M}^\alpha  (z,p) =2 p^\alpha  {\cal M}_p (-(zp), -z^2) + z^\alpha  {\cal M}_z (-(zp),-z^2)
\ . 
\end{align}
When  $z^2 \to 0$,
the ${\cal M}_p $ part gives the twist-2 distribution,  
while $ {\cal M}_z $ is a  purely higher-twist contamination. 
Note, 
however,  that $ {\cal M}_z $  does not  contribute
 in the standard 
definition of  unpolarized TMDs  because
it is  based on taking $z=(z_-, z_\perp)$ in the $\alpha=+$  component of 
${\cal O}^\alpha$.  Then the   \mbox{$z^\alpha$-part}  drops out.
As  a result,   
   ${\cal F}(x, k_\perp^2)$   is 
 related to  $ {\cal M}_p (\nu, z_\perp^2)$  by the scalar formula 
(\ref{MTMD}).  

To keep a simple relation between TMDs and \mbox{$z_3$-pseudo-PDFs} 
(and, hence, quasi-PDFs), we need to   arrange that 
 the $z^\alpha$ contamination does not contribute to pseudo-PDFs. This is achieved by    taking  the
  time component of \mbox{$ {M}^\alpha  (z=z_3,p)$}  and defining 
   \begin{align}
 { M}^0   (z_3,p)= & 2 p^0    {\cal M}_p   (\nu,z_3^2)   
  =   2 p^0   
 \int_{-1}^1 dx \,  
{\cal P} (x,  z_3^2)   \, 
 \,  e^{i  x\nu   }  \  . 
 \label{MpP}
\end{align} 
The quasidistribution $Q(y,P) $
is defined in a similar way:
 \begin{align}
{ M}^0   (z_3,p)= & 2 p^0  
 \int_{-\infty}^\infty dy\,  
Q(y,P) \, 
 \,  e^{i  y Pz_3 }  \  . 
 \label{OPhixspin12}
\end{align}

As a result,   the  connection between $Q(y,P)$   and 
 ${\cal P} (x,  z_3^2) $ is given 
by the scalar  formula (\ref{Py}). 
From now on, we will  shorten the notation by using  $ {\cal M}_p \to  {\cal M}$.

In QCD,  $ {\cal M} (\nu,  z_\perp^2) $ contains  \mbox{$\sim \ln z_\perp^2$}  terms 
corresponding to   the  $\sim 1/k_\perp^2$ hard tail of ${\cal  F} (x, k_\perp^2)$.
Thus,  it makes sense to visualize 
$ {\cal M}  (\nu, z_\perp^2)  $    
as a sum of a soft part $ {\cal M}^{\rm soft} (\nu, z_\perp^2)$, 
that has a finite $z_\perp^2\to 0$ limit,  
and a  logarithmically singular {\it hard part}      reflecting   the evolution. 
The same applies to   $ {\cal M}  (\nu, z_3^2)  $.

 \setcounter{equation}{0}

\section{Hard contribution in coordinate space}

      Even if one starts with a purely soft  TMD (or pseudo-PDF),   
 the      gluon exchanges
 generate the hard part.   
Our goal is  to 
 describe  this  on the  operator level,
as a modification of the original 
soft  %%%%%%%%%%%
 bilocal operator by 
 gluon corrections.

 \subsection{Link self-energy  contribution}

To begin with, we consider the modification resulting from the 
self-energy correction  to  the  gauge 
 link    (see also Refs.  \mbox{\cite{Polyakov:1980ca,Chen:2016fxx}).} 
 At one loop, it is   given by
 \begin{align} 
 \Gamma_ \Sigma (z)  = &  ({i} g)^2\,C_F \,\frac12 \,  \int_0^1 dt_1 \,  \int_0^1 dt_2 \, 
  z^\mu z^\nu \,  D_{\mu \nu} ^c [ z (t_2-t_1) ] \  , 
    \label{self}
 \end{align}
 where  $D_{\mu \nu} ^c[z (t_2-t_1)]$ is the gluon propagator connecting points
 $t_1 z$ and $t_2 z$.  For massless gluons  in Feynman gauge, we have 
  $D_{\mu \nu} ^c (z)  =- g_{\mu \nu}  /4\pi^2 z^2$, and end  up with a divergent integral
    \begin{align} 
     \int_0^1 dt_1 \,  \int_0^1 \frac{dt_2}{(t_2-t_1)^2}  \,  .
    \label{t1t2}
 \end{align}
 This divergence has an ultraviolet origin. 
 For spacelike $z$, it may be regularized by  
 using the Polyakov prescription
 \cite{Polyakov:1980ca} \mbox{$1/z^2 \to 1/(z^2-a^2)$}  for the gluon propagator. 
 Taking $z=z_3$  we have
  \begin{align} 
  \Gamma_ \Sigma (z_3,a)  = & - g^2\,C_F \,\frac{z_3^2}{8 \pi^2}
   \,  \int_0^1 dt_1 \,  \int_0^1  \, \frac{dt_2}{ z_3^2 (t_2-t_1)^2 + a^2}  \  . 
    \label{selfa}
 \end{align}
  Calculating the  integrals gives the result 
  \begin{align} 
 \Gamma_  \Sigma (z_3,a)  =   -\,C_F \,\frac{\alpha_s}{2 \pi}
  & 
  \left [ 
   \,
 2  \frac{ |z_3|}{a} \,  \tan
   ^{-1}\left(\frac{|z_3|}{a}\right)   -  \ln 
   \left(1+ \frac{z_3^2}{a^2}\right) \right ] \  
       \label{selfex}
 \end{align}
 coinciding with that given  in Ref. \cite{Chen:2016fxx}.

 If we keep $z_3$ fixed and take the  small-$a$ limit, we can expand
   \begin{align} 
   \Gamma_\Sigma (z_3,a) |_{a \to 0}  = &  -\,C_F \,\frac{\alpha_s}{2 \pi}
  \left [
   \,
\frac{\pi |z_3|}{a}  - 2 -  \ln \frac{z_3^2}{a^2}  + {\cal O}(a^2/z_3^2) \right ] \   
       \label{selfexexp}
 \end{align}
 (see also Ref. \cite{Ishikawa:2017faj}).  
  This result clearly shows a linear divergence $\sim |z_3|/a$  in the $a \to 0$ limit.
 As explained in the pioneering paper  \cite{Polyakov:1980ca}, it may be interpreted as the mass renormalization
 $\delta m  $ of a test particle moving along the link,
     \begin{align} 
\delta m  =   \, C_F \,\frac{\alpha_s}{2 \pi} \, \frac{\pi}{a} \ . 
       \label{delta_m}
 \end{align}

 Alternatively, for a  fixed $a$ and small $z_3$, the factor in the square brackets behaves 
 like $z_3^2/a^2$, i.e., $  \Sigma (z_3,a) $  vanishes at $z_3=0$. 
  This  distinctive feature of  such  UV contributions 
   is a mere consequence of the fact that 
  the gauge link converts into unity for $z_3=0$. 
  
 As a result,  the  UV terms vanish on the light cone,
 and that is why they are usually not  discussed  in the context of  the light-cone  PDFs.
 Also, the fact that     $\Sigma (z_3=0,a) =0$  means that, at fixed $a$,  $\Sigma$  gives 
 no corrections to the vector current, i.e.  the  number of  the  valence quarks  is not changed.

   \begin{figure}[t]
   \centerline{\includegraphics[width=2in]{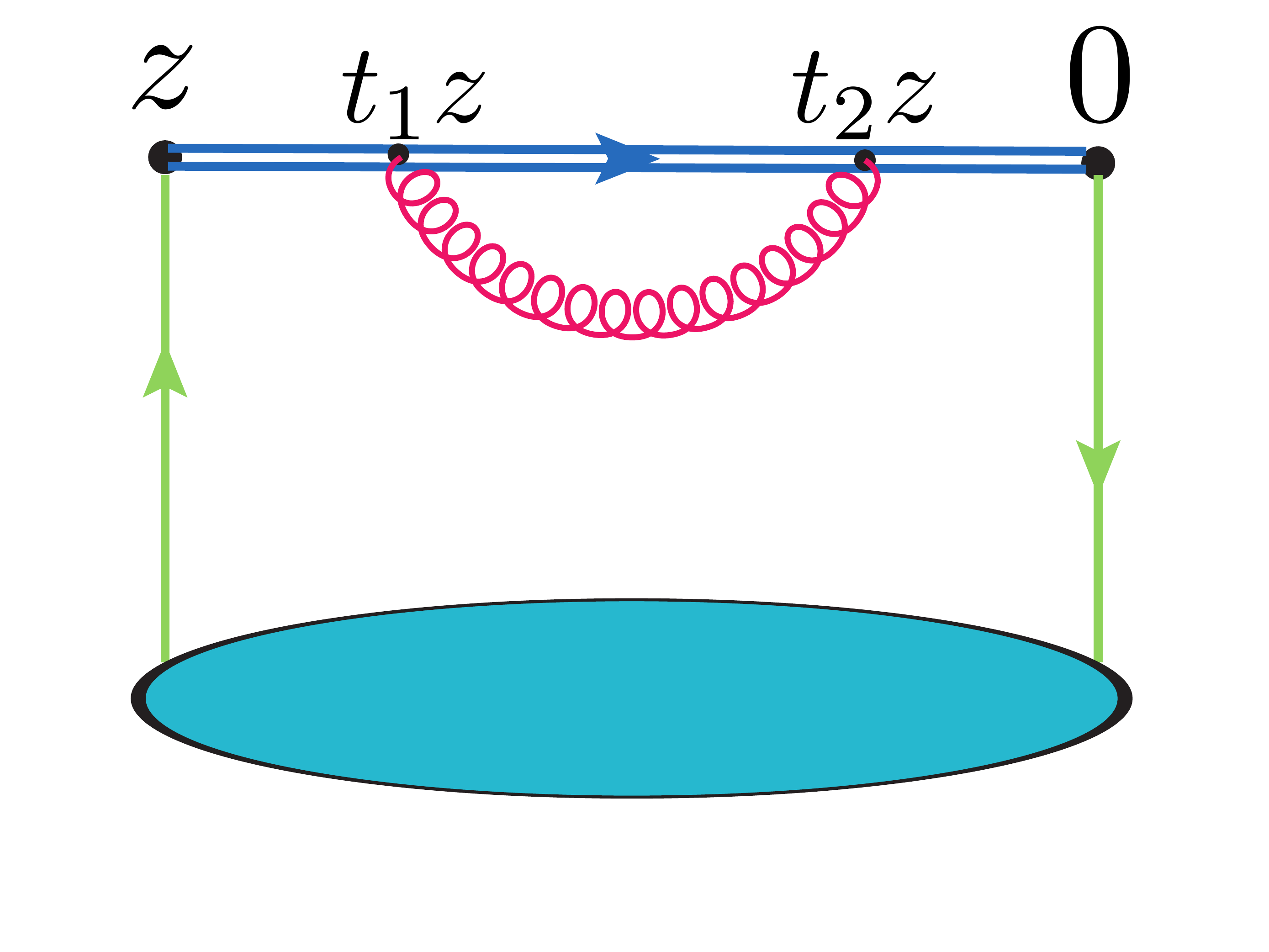}}
   \caption{Renormalization  of the gauge link.
   \label{linkself}}
   \end{figure}
 
 Eq. (\ref{selfexexp})  also shows a logarithmic divergence corresponding to the  anomalous dimension 
 $2\gamma_{\rm end}$ due to two  end-points of the link. In  the lowest order  (see, e.g. \cite{Aoyama:1981ev})
    \begin{align} 
\gamma_{\rm end} =   -\,C_F \,\frac{\alpha_s}{4 \pi} \  . 
       \label{gamma_end}
 \end{align}
 
  In an Abelian theory, the vacuum average of an exponential 
 is the exponential of the one-loop vacuum average.
 Hence, the one-loop  term exponentiates.  In the $a\to 0$ limit, this 
    produces  a factor
    \begin{align} 
Z_{\rm link} (z_3,a)= e^{-\delta m |z_3|} e^{ - 2 \gamma_{\rm end} \ln (z_3^2/a^2)}  \  . 
       \label{Z}
 \end{align}
 The exponentiation works also in a non-Abelian case 
 \cite{Gatheral:1983cz,Frenkel:1984pz,Korchemsky:1987wg}, 
 but the exponential involves then   higher-$\alpha_s$ corrections accompanied 
 by higher  irreducible color factors.  The all-order renormalization of Wilson loops 
 and lines was discussed in Refs. \cite{Dotsenko:1979wb,Brandt:1981kf,Aoyama:1981ev}.

The Polyakov prescription \mbox{$1/z_3^2 \to 1/(z_3^2+a^2)$} 
softens the gluon propagator at distances $z_3\sim $ several $a$,  
and eliminates its singularity at $z_3=0$. In this respect,
it is similar to  the UV regularization  produced by a  finite lattice spacing. 
Thus, we find it instructive to use this prescription  in the studies of the UV properties
of pseudo-PDFs.

\subsection{Vertex contribution}

Working in Feynman gauge, we need also to consider   vertex diagrams
involving gluons  that connect the gauge link with the quarks, see  Fig. \ref{link}.

 \subsubsection{Basic structure}

\begin{figure}[t]
   \centerline{\includegraphics[width=2.5in]{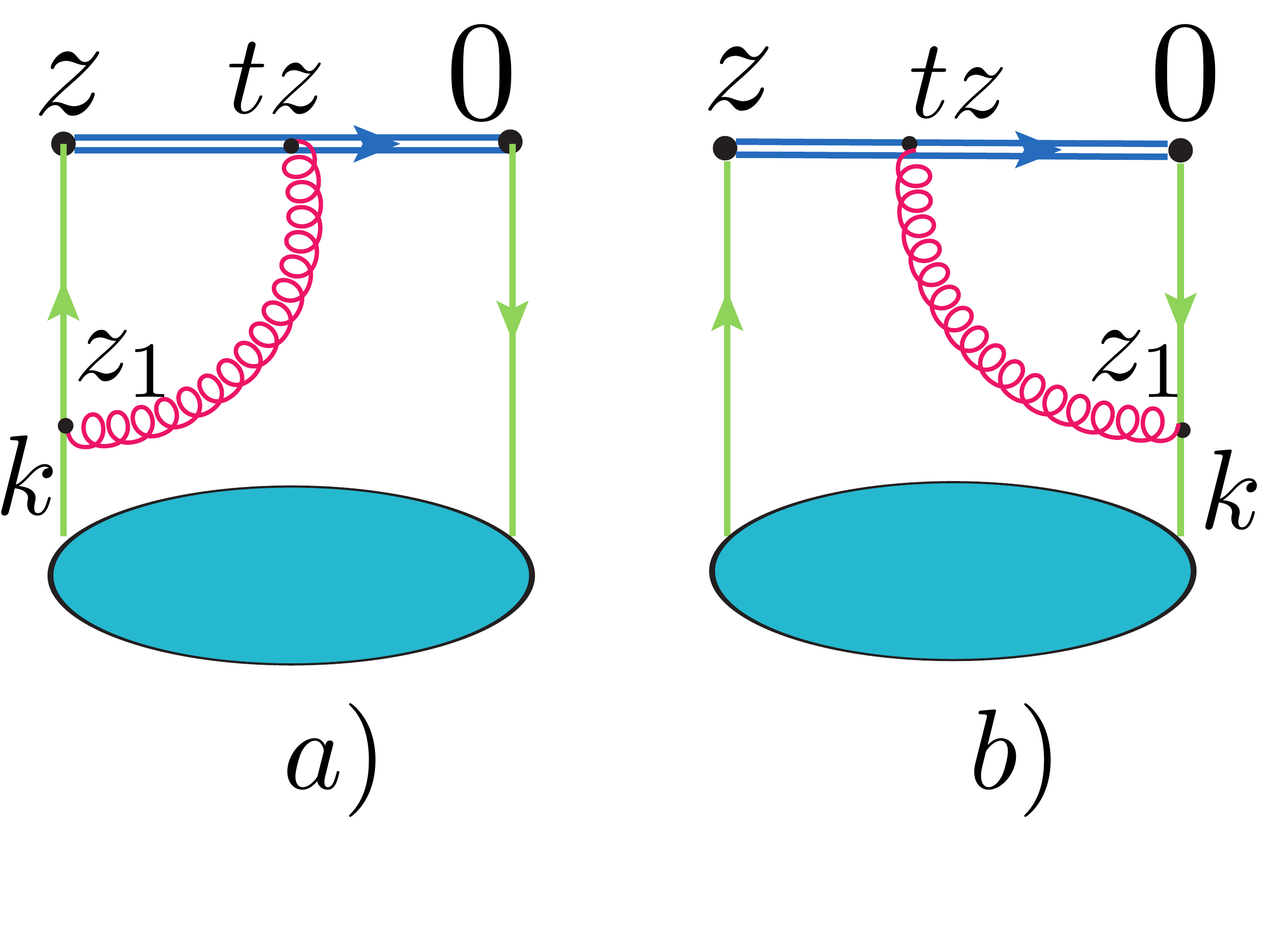}}
        \vspace{-10mm}
   \caption{Insertions of gluons coming out of the gauge link.
   \label{link}}
   \end{figure}

There are two possibilities: gluon may be connected to  the left  
(Fig. \ref{link}a) or to the right  (Fig. \ref{link}b) 
quark leg.  To facilitate integration over $z_1$, we write  the fields at this point 
 in the momentum representation.
If  the gluon  is inserted into the right quark  line, we start with 
   \begin{align} 
  O_R^\alpha   (z) = &  (i g)^2\,C_F \, \int_0^1 dt \, 
 \int   d^4z_1   \,D^c (z_1-tz )   
  \nonumber \\ &  \times   \int d^4 k \, e^{i (kz_1)}    \,    \bar \Psi (k)  \slashed z  S^c(z_1) \gamma^\alpha  \psi (z)   ] \  . 
    \label{gauge1}
 \end{align}
The insertion into the left    leg gives a similar expression. 
 Using $S^c (z) = i\slashed z/ 2\pi^2 (z^2)^2$ and $D^c(z) = 1/4\pi^2 z^2$ combined with the
representation
    \begin{align} 
\frac1{(z^2/4)^{N+1}} = \frac{i^{N-1}}{ N!} \int_0^\infty d \sigma \sigma^N e^{-i \sigma z^2/4} 
\label{z2N}
 \end{align}
gives a Gaussian integral   over  $z_1$. Taking it, we obtain    
 \begin{align} 
  O^\alpha_R  (z) = &i     \frac{ g^2}{8 \pi^2} \, C_F  \int_0^1 dt \, 
 \int_0^\infty \sigma_1 \frac{d\sigma_1 d \sigma_2 }{(\sigma_1 + \sigma_2)^3}
   \nonumber \\ &  \times \int d^4 k \,  e^{i  (t (k z)  \sigma_2 -
    \sigma_1  \sigma_2 t^2 z^2/4 )/(\sigma_1 + \sigma_2)} 
  \nonumber \\ &  \times  
 \,  \bar \Psi (k)   \left  [ (kz)  
   + t \sigma_2  z^2/4  \right ]   \gamma^\alpha  \psi (z)   
   \ .
  \label{RightV}
 \end{align}

 The external quark lines enter into the soft part, 
 thus we neglect their virtuality, i.e. 
 the $k^2$ term in the exponential. 
For the same reason, we   also   neglect  the $\bar \Psi (k) \slashed k$ term.  

Switching to $\lambda_i=1/\sigma_i$,   introducing common 
\mbox{$\lambda=\lambda_1+\lambda_2$,}   with $\lambda_1/\lambda =\beta$,
then relabeling \mbox{$\lambda=1/\sigma$,}
we obtain  
 \begin{align} 
&     O^\alpha_{R}  (z)  = i    \,      \frac{ g^2}{8 \pi^2} \, C_F  \int_0^1 dt \, 
    \int_0^\infty {d\sigma}\,    e^{-  i \sigma
    t^2 z^2/4} 
    \int_0^1 d \beta \,
  \nonumber \\ &  \times 
   \int d^4 k \,      e^{i \beta t (kz)} 
 \,  \bar \Psi (k)  
 [(1-\beta) (kz)/ \sigma  +t\, z^2/4] \, 
     \gamma^\alpha  \psi (z)
   \ .
    \label{RightVI}
 \end{align}

 \subsubsection{UV singular   term}
 
We can take the $d^4k$ integral for 
 the   second, namely  $t z^2$,  term  in Eq. (\ref{RightVI})   to  get 
 \begin{align} 
&     O^\alpha_{R2}  (z)  = i    \,      \frac{ g^2}{8 \pi^2} \, C_F  \int_0^1 dt \, t 
    \int_0^1 d \beta \,
  \nonumber \\ &  \times   \frac{ z^2}{4}    \int_0^\infty {d\sigma}\,    e^{-  i \sigma
    t^2 z^2/4}  \, 
 \,  \bar \psi (t\beta z)  
\, 
     \gamma^\alpha  \psi (z)
   \ .
    \label{RightV2}
 \end{align}
 
Now, integration over $\sigma$
leads to  an UV divergence from the small-$t$ integration.
 The situation is similar to that  in the case of  the  link renormalization.
 Thus, we regularize 
 $1/z^2 \to 1/(z^2-a^2)$ in the initial expression (\ref{z2N}) for the gluon propagator. 
 Since  the singularity is accompanied by the quark field 
 $ \bar  \psi (t \beta z)  $ at  $t=0$, we 
 isolate it by splitting   $\bar  \psi (t \beta z) $  into 
$ \bar  \psi (0)
 + \left [ \bar  \psi (t \beta z)  -  \bar  \psi (0) \right ]
$. The UV-singular term is then given by 
  \begin{align} 
  O^\alpha_ {\rm R, UV}  (z,a) = &  \frac{ g^2}{8 \pi^2} \, C_F \, \bar \psi (0)  \gamma^\alpha  \psi (z)    
  \nonumber \\ &  \times  
   \int_0^1 d \beta \,   \int_0^1 dt\, 
  \frac{t z^2 }{ t ^2z^2-  a^2/(1- \beta) } \,  . 
    \label{UVaR}
 \end{align}
Taking integrals over $t$ and $\beta$    gives the expression 
    \begin{align} 
 O^\alpha_ {\rm R, UV}  (z,a) = &  \frac{ g^2}{16 \pi^2} \, C_F \, \bar \psi (0)  \gamma^\alpha  \psi (z)    \nn  &
  \times \left [  \left (1-\frac{a^2}{z^2}  \right  )  \, 
  \ln  \left (1- \frac{z^2}{a^2} \right  )  -1\right ]
       \  
    \label{UVsing}
 \end{align}
that   contains   the same $ \ln  \left (1- {z^2}/{a^2} \right  ) $  logarithmic  term 
as in the self-energy correction (\ref{selfex}).
In the $a\to 0$ limit, this result agrees with that obtained in Ref. \cite{Ishikawa:2017faj}.
The diagram with  the insertion into  the left leg gives the same contribution, thus the total UV-singular contribution doubles
$
  O^\alpha_ { \rm UV}  (z,a) = 2  O^\alpha_ {\rm R, UV}  (z,a)$.

Switching  to  $z=z_3$, we have  the $\ln (z_3^2/a^2)$ structure  in the $a\to 0$ limit, 
 and  we  may 
  combine it  with the UV divergences generated by the link self-energy diagrams.
Again, for a fixed $a$,  the $ O^\alpha_ {\rm UV}  (z_3,a) $ 
contribution vanishes in the $z_3^2 \to 0$ limit.

 \subsubsection{UV finite  term}

The $ \left [ \bar  \psi (t \beta z)  -  \bar  \psi (0) \right ]$ term vanishes for $t=0$, and for this reason its  contribution
    \begin{align} 
   O^\alpha_{\rm R,reg}  (z_3,a) = &    \,        \frac{ g^2}{8 \pi^2} \, C_F\, z_3^2
\int_0^1  dt \, t    
   \int_t^1 \, \frac{  d \beta }{t^2z_3^2 +\beta^2 a^2/(1-\beta) } \,    \nonumber \\ &  \times 
   [ \bar \psi (t z_3 )    \gamma^\alpha  \psi (z_3) - \bar \psi (0 )  \gamma^\alpha  \psi (z_3)]
     \
    \label{UVareg}
 \end{align}
 is finite in the $a\to 0$ limit.   Taking $a=0$  
  and integrating over $\beta$  gives 
  \begin{align} 
  O^\alpha_{\rm R,reg}  (z_3, a=0) = &    \,     \frac{ \alpha_s}{2\pi} \, C_F
 \,\int_0^1  du \left[ \frac{\bar u }{u }\right ] _+ 
  \bar \psi (uz_3) \gamma^\alpha \psi (z_3)
     \ ,
    \label{UVRReg}
 \end{align}
 where we  have relabeled $t \to u$, and  introduced  $\bar u \equiv 1-u$. The plus-prescription is defined by
   \begin{align} 
 \,\int_0^1  du \left[ \frac{\bar u }{u }\right ] _+  F(u)  =    \,\int_0^1  du \, \frac{\bar u }{u } \,  [F(u)-F[0] \ ,
    \label{Plus}
 \end{align}
 assuming that $F(0)$ is finite.
Again, the plus-prescription structure  of Eq. (\ref{UVRReg})
guarantees that this term   gives no corrections to the local current.
 
Adding  the contribution of the diagram with the left-leg insertion, we may write
the total regular term as 
    \begin{align} 
  O^\alpha_{\rm reg}  (z_3, a=0) = &    \,     \frac{ \alpha_s}{2\pi} \, C_F
 \,\int_0^1  du \int_0^1  d\varv \, \bar \psi (uz_3) \gamma^\alpha \psi (\bar \varv z_3) 
  \nn & \times  \left \{ \delta (\varv) \left[ \frac{\bar u }{u }\right ] _+
 + \delta (u) \left[ \frac{\bar \varv }{\varv }\right ] _+   \right \}  % 
     \ .
    \label{gauge24}
 \end{align}
 
 In terminology of Balitsky and Braun  \cite{Balitsky:1987bk},   $\bar \psi (uz) \gamma^\alpha \psi (\bar \varv z) $
 is a ``string operator''.  It involves two fields on the straight line segment  $(0,z)$,  separated 
from its endpoints  by $uz$ and   $\varv z$, respectively. The difference with  Ref. \cite{Balitsky:1987bk} is that  
 $z$  here   is not  on the light cone $z^2=0$.
 
  \subsubsection{Evolution  terms}
 
Now, let us analyze  the $(1-\beta) (kz)/ \sigma$    term  in Eq. (\ref{RightVI}),   
  \begin{align} 
  O^\alpha_{\rm R, Evol}  (z)  =& i    \,      \frac{ g^2}{8 \pi^2} \, C_F  \int_0^1 dt \, 
    \int_0^\infty \frac{d\sigma}{\sigma}\, e^{-  i \sigma
    t^2 z^2/4}   \int_0^1 d \beta \, (1-\beta)  \nonumber \\ &  \times 
     \int d^4 k \,    (kz)\, 
     e^{i  t \beta  (k z)} 
   \bar \Psi (k)    \gamma^\alpha \psi (z)
   \ .
    \label{LogIn}
 \end{align}
 This expression   has  
 a   logarithmic    infrared divergence  resulting from the 
 lower limit of the $\sigma $ integration.
 
 In fact, if we would keep the $k^2$ term in the exponential 
 of Eq.  (\ref{RightV}), it would produce a  $e^{ik^2/\sigma}$-type factor
 in the expression above, and there would be no infrared divergence. 
 In other words, the quark  virtuality would provide an IR cut-off. 
 In the coordinate representation, this corresponds to a cut-off 
 produced by the nonperturbative $z_1^2$-dependence of the 
 soft matrix element  $\langle p|\bar \psi (0) \gamma \psi (z_1) |p \rangle$ 
 (we refer here to Fig.\ref{link} a).
 
 An example of calculations, in which such a dependence was taken into account,
 may be found in our paper \cite{Radyushkin:2015gpa}, where the virtuality 
 distribution formalism was applied to the pion  transition form factor. 
 In our  present  context, the cut-off will be provided by 
 the $k_\perp$-dependence of the soft part of the TMD ${\cal  F} (x, k_\perp^2)$,
 i.e. by  the hadron size. 
 Leaving  a detailed investigation for future studies, we 
 will just  assume now some reasonable form   of  an IR cut-off function.
 
In  particular, the IR singularity may be  regularized  by the 
  $e^{-i m^2/\sigma}$ factor,
which is equivalent  to  adding the same mass term $m^2$ to both  propagators. 
In this case (switching to $z^2=-z_3^2$),  we define 
   \begin{align} 
    L_m (t^2z_3^2 m^2)  \equiv 
 \int_0^\infty    \frac{d \sigma }{ \sigma }\,  e^{  i \sigma  t^2  z_3^2/4-i m^2/\sigma} =
 2 K_0 (t m |z_3|)  \   . 
    \label{Lzm}
 \end{align}
  This function has  a logarithmic $\ln z_3^2 m^2$ singularity
  for small $z_3$ and an exponential  $e^{-|z_3|m} $ fall-off  for large $z_3$.
  One should realize that since $m$ parametrizes 
  the IR cut-off imposed by the hadron size, numerically $m$  
  should be of an order of 0.5 GeV.
  
 Another simple possibility  to regularize the IR 
 singularity is to cut the $\sigma$-integral from below. Then we have 
    \begin{align} 
    L_{1/z_0}  (t^2z_3^2 /z_0^2)  \equiv 
 \int_{1/z_0^2}^\infty    \frac{d \sigma }{ \sigma }\,  e^{  i \sigma  t^2  z_3^2/4} =
 \Gamma[0, t^2z_3^2/4 z_0^2 ] \   ,
    \label{LzL}
 \end{align}
 where $\Gamma[0,x]$ is the incomplete gamma-function.
 Again, we have   a logarithmic $\ln z_3^2 /z_0^2$ singularity
  for small $z_3$, but  now  
 a Gaussian $e^{-z_3^2 /4z_0^2}$ behavior for large $z_3$.
 As observed in \mbox{Ref. \cite{Radyushkin:2015gpa},}
 the two forms of the IR cut-off automatically 
 follow from   soft distributions with the exponential and Gaussian 
 fall-off at large $z_\perp$, respectively.

 For small $|z_3|$, we may write   in both cases  
    \begin{align} 
  L_ \Lambda  (t^2 z_3^2 \Lambda^2)=  L_ \Lambda ( z_3^2 \Lambda^2) - 2  \ln t + {\cal O} (\Lambda^2 z_3^2) \  . 
  \label{k0} 
   \end{align}
The logarithms  $\ln (z_3^2  \Lambda^2)$  contained in  $ L_\Lambda  ( z_3^2  \Lambda^2)$ 
 reflect  the perturbative  evolution.  

Integrating over $t$ in the part corresponding to the $ L_ \Lambda  ( z_3^2  \Lambda^2) $  
term in Eq. (\ref{k0}), 
 changing the  notation $\beta \to \varv$, and adding the contribution from the left-leg insertion,  we get 
     the total  logarithmic  contribution  coinciding with  that of Ref.  \cite{Balitsky:1987bk} 
    \begin{align} 
  O^\alpha_{\rm  log}  (z_3) & =     \,    L_ \Lambda  ( z_3^2  \Lambda^2) \,   
   \frac{ \alpha_s}{2\pi} \, C_F
 \,\int_0^1  du \int_0^1  d\varv \nn & \times  \, \left \{ \delta (u) \left[ \frac{\bar \varv }{\varv }\right ] _+
 + \delta (\varv) \left[ \frac{\bar u }{u }\right ] _+   \right \} \, \bar \psi (u z_3) \gamma^\alpha \psi (\bar \varv z_3) 
     \ .
    \label{Logtot}
 \end{align}
  For small $z_3^2$, this  result corresponds to the following correction to the ITD
    \begin{align} 
  {\cal M}_{\rm  log}  (\nu,  z_3^2) & =     \,    L_ \Lambda  ( z_3^2  \Lambda^2) \,    \frac{ \alpha_s}{2\pi} \, C_F
 \,\int_0^1  dw 
 \left[ \frac{2 w }{1-w }\right ] _+  \,      {\cal M}^{\rm soft}   (w \nu,  0)  \ .
    \label{LogtotM}
 \end{align}

Note  that, in contrast to  the UV divergent contribution,
the  $L_ \Lambda(z_3^2  \Lambda^2)$ function is singular in the $z_3^2\to 0$ limit, and 
the parameter 
 $|z_3|$  in the integrals  of Eqs. (\ref{Lzm}), (\ref{LzL})  works like an  ultraviolet 
 rather than an  infra-red cut-off.

 \subsubsection{IR finite  term}

The  $\ln t$-term in Eq. (\ref{k0})  produces an IR finite contribution 
     \begin{align} 
  O^\alpha_{\rm R, Fin}  (z)  =&-  i    \,      \frac{ g^2}{4 \pi^2} \, C_F  \int_0^1 dt \, \ln  t \, 
    \int_0^1 d \beta \, (1-\beta)       \nonumber \\ &  \times    \int d^4 k \,    (kz)\, 
     e^{i  t \beta  (k z)} 
     \,    \bar \Psi (k)    \gamma^\alpha \psi (z)
   \ .
    \label{Lnt}
 \end{align}

Transforming the $\beta$-integral through integrating exponential   by parts,  
changing  $\beta=\varv/t$  and adding the  left $O^\alpha_{\rm L, Fin}  (z)$ contribution, we obtain 
   \begin{align} 
  O^\alpha_{\rm  Fin}  (z)  =&-     \,      \frac{ \alpha_s }{ \pi} \, C_F  \int_0^1 du \int_0^1 d\varv \, 
\bar \psi (\varv z)    \gamma^\alpha \psi (\bar u z) 
       \nonumber \\ &  \times 
   \left [  \delta (u) \,  s_+(\varv)+ \delta (\varv) \,  s_+(u) \right ]
   \ , 
    \label{Ftot}
 \end{align} 
 where $s_+(u)$ is the plus-prescripted version of $s(u)$ given by 
    \begin{align} 
s(u) & \equiv 
  \int_u^1 dt \,\frac{ \ln t}{t^2}  = \frac{1-u+\log (u)}{u}  \   . 
    \label{Sv}
 \end{align}
Since this part   depends on $z$   through the  fields only,  we deal  with  a finite radiative correction
to the soft contribution. 

\subsection{Quark-gluon exchange contribution}

\begin{figure}[t]
   \centerline{\includegraphics[width=2.5in]{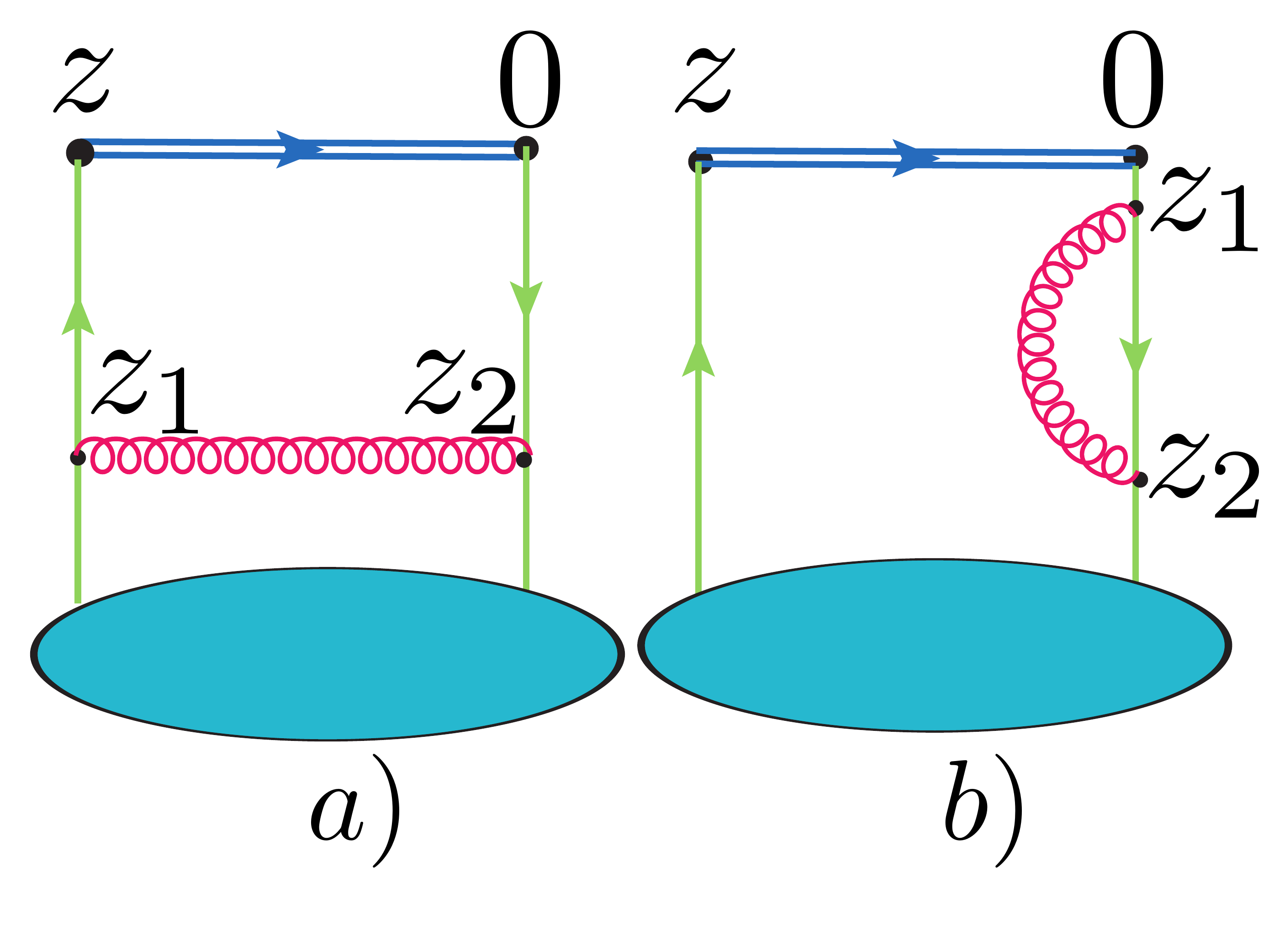}}
        \vspace{-5mm}
   \caption  {a)  Gluon  exchange diagram.  b)  One of quark self-energy correction diagrams.
   \label{quarkself}}
   \end{figure}

There is also a 
contribution to the hard part  given by the diagram  \ref{quarkself}a  
containing  a gluon exchange between
two quark lines.  Taking  the time component $\alpha=0$, we have
    \begin{align} 
  O^0_{\rm  exch}  (z_3) & =     \,      \frac{ \alpha_s}{2\pi} \, C_F
 \,\int_0^1  du \int_0^{1-u}   d\varv \nn & \times  \, \left \{ L_R  ( z_3^2 R^2) \,  - 1  \right \} \, \bar \psi (u z_3) \gamma^0 \psi (\bar \varv z_3) 
     \ .
    \label{exch}
 \end{align}
For the ITD, this gives  the integral 
    \begin{align} 
  {\cal M}_{\rm  exch}  (\nu,  z_3^2) & =     \,       \frac{ \alpha_s}{2\pi} \, C_F
 \,\int_0^1  dw  \, (1-w) 
    \nn & \times  \, 
 \left\{ L_R  ( z_3^2 R^2) - 1   \right \}  \,      {\cal M}^{\rm soft}   (w \nu,  0)  \ 
    \label{exchM}
 \end{align}
with   the  \mbox{$(1-w)$}   integrand.   Its logarithmic part, combined with the $2w/(1-w)$
term coming 
from the vertex correction,  gives the 
expected form $(1+w^2)/(1-w)$   \cite{Altarelli:1977zs}   of the evolution kernel.
Note, however, that  unlike the vertex part, the exchange contribution (\ref{exchM})
does not have the plus-prescription form.

 The standard  expectation  is that
one would  get it after the addition of the 
  quark self-energy diagrams,  one of which is shown in Fig. \ref{quarkself}b.
  As   usual, one should take just a half 
of each,  absorbing  the  other halves into the  soft part.
These diagrams have an 
 ultraviolet divergence that  may be regularized  by  the same (for uniformity) 
 Polyakov prescription  $1/z^2\to 1/(z^2 -a^2)$ for the gluon propagator. The result is a 
$\ln (a^2 m^2)$  contribution.  But, since  it has no $z$-dependence,
it cannot help one  to get the  plus-prescription form for the 
logarithmic (in $z_3^2$)  part of the exchange contribution.

A possible way out is to represent  $\ln (a^2 m^2)$   as   the difference $\ln (z_3^2 m^2)   -  \ln (z_3^2 /a^2) $
of the evolution-type logarithm $\ln (z_3^2 m^2)  $ and a UV-type logarithm  $ \ln (z_3^2 /a^2) $.
The latter  can be  added 
to the UV divergences of the diagrams \ref{linkself} and \ref{link},
so that the total UV divergent contribution 
is   
 \begin{align} 
\Gamma_{\rm UV} (z_3, a)   =   - \,\frac{\alpha_s}{2\pi}\, C_F
   & \left [  
   \,
 2  \frac{ |z_3|}{a} \,  \tan
   ^{-1}\left(\frac{|z_3|}{a}\right)   \right. \nn & \left.  - 2  \,  \ln 
   \left(1+ \frac{z_3^2}{a^2}\right) +\frac12 \,  \ln 
   \left( \frac{z_3^2}{a^2}\right) \right ] \,  . 
       \label{GUV}
 \end{align}
The $\ln (z_3^2 m^2)$  part  should be added to the evolution  kernel, and it converts the $(1-w)$ term into 
$(1-w)_+$.

\subsection{Reduced Ioffe-time distribution}

Another possibility is to use 
the reduced Ioffe-time distribution  of Refs. \cite{Radyushkin:2017cyf,Radyushkin:2017sfi,Orginos:2017kos,Karpie:2017bzm} 
 \begin{align}
{\mathfrak M} (\nu, z_3^2) \equiv \frac{ {\cal M} (\nu, z_3^2)}{{\cal M} (0, z_3^2)} \  . 
 \label{redm0}
\end{align}
Then the  UV divergences generated by the link-related  and quark-self-energy diagrams 
cancel in the ratio (\ref{redm0}).\footnote{According to Ref. \cite{Ishikawa:2017faj}, 
the cancellation holds to all orders in perturbation theory.  See also \cite{Ji:2017oey,Green:2017xeu} ,
and earlier papers   \cite{Dotsenko:1979wb,Brandt:1981kf,Aoyama:1981ev,Craigie:1980qs}.}
 Furthermore,
since $\nu=0$ is equivalent to $p=0$, the denominator factor 
automatically completes 
the gluon-exchange contribution $(1-w)$ to $(1-w)_+$.

The vertex part (\ref{LogtotM}) of the evolution 
 kernel  has  the plus-prescription 
 structure from the start. For this reason, it  does  not contribute to  the denominator factor 
 ${\cal M} (0, z_3^2)$. 
As a result,   ${\mathfrak M} (\nu, z_3^2)  $ satisfies the evolution equation 
    \begin{align}
    \frac{d}{d \ln z_3^2} \,  
{\mathfrak M} (\nu, z_3^2)    &= - \frac{\alpha_s}{2\pi} \, C_F
\int_0^1  dw \,   B ( w ) \,   {\mathfrak  M} (w \nu, z_3^2)  
\label{EE}
 \end{align}
with respect to
$z_3^2$,  where $B(w)$ is the 
 full plus-prescription  Altarelli-Parisi  (AP) evolution kernel \cite{Altarelli:1977zs} 
 \begin{align} 
B(w)  =&
       \left [\frac{1+w^2} {1-w}   \right ]_+
  \ .
     \label{V1}
  \end{align}   

There are also  non-logarithmic terms\footnote{Such terms also appear
in the expression for the pseudo-PDF in  Ref. \cite{Ji:2017rah}. 
There is a difference  with  %%%%%%%%%%%%%
our results, because its authors use dimensional regularization of IR singularities.}%%%%%%%%%%%
\   from (\ref{gauge24}) and   (\ref{Ftot})
that contribute to the  numerator factor  ITD ${\cal M} (\nu, z_3^2)$. 
However, since they have the plus-prescription 
 structure,   they vanish in   the denominator factor 
 ${\cal M} (0, z_3^2)$. 
Thus,  we get the one-loop expression for the hard part of  the reduced 
ITD in the following form 
 \begin{align} 
{\mathfrak M}^{\rm hard}  (\nu, z_3^2)    =&  -  \frac{\alpha_s}{2\pi} \, C_F
\int_0^1  dw \,  \left \{   \left (\frac{1+w^2} {1-w}   \right )_+   \left [ \ln ( z_3^2m^2 e^{2\gamma_E}/4)  +1 \right ]
\right. \nn   & \left. 
+ 4   \left [\frac{\log (1-w)}{1-w} \right]_+ 
\right \} 
{\mathfrak  M}^{\rm soft}  (w \nu, 0)   \  , 
\label{Mha}
 \end{align}
 where we also have explicitly displayed the 
 logarithmic part of the modified Bessel function 
$K_0 (|z_3| m)$.

 \setcounter{equation}{0}
 
  \section{Hard contribution to quasi-PDFs}

 The Ioffe-time distributions are basic starting objects 
 for all parton distributions. Hence, the results obtained above
 may be used to calculate the one-loop corrections for quasi-PDFs.
 In what follows, we analyze how the $z_3^2$-dependence 
 of the one-loop hard part 
  of   ${\cal  M}^{\rm hard}  (\nu, z_3^2)$ is reflected  
  in some specific features of quasi-PDFs.
  For the soft part, we will assume the collinear approximation   %%%%%%%%
  ${\cal M}^{\rm soft}  ( \nu, z_3^2) = {\cal M}^{\rm soft}  ( \nu, 0)$. %%%%%

\subsection{Ultraviolet divergent  terms}

 For the UV-singular terms, we have 
  \begin{align} 
{\cal  M}_{\rm UV}  (\nu, z_3^2)    =& 
\Gamma_{\rm UV}(z_3,a) \,   {\cal M}^{\rm soft}  ( \nu, 0)  \  . 
\label{MUVa}
 \end{align}
 
Combining the definition (\ref{QyP}) of the quasi-PDFs 
with the relation (\ref{fxMnu})  between  $ {\cal  M}   ( \nu, 0)$
and  the PDF $f(x)$ we have
  \begin{align}
     Q_{\rm UV}(y,P) =  \int_{-1}^1 dx\,
R_{\rm UV}  (y-x;a) \, f(x) \  , 
  \label{QUVa}
\end{align} 
where 
  \begin{align}
  R_{\rm UV}  (y-x;a)   =\frac{P}{2 \pi}   \int_{-\infty}^\infty dz_3\,
      e^{-i (y-x)  Pz_3}\, \Gamma_{\rm UV}(z_3,a) \, 
 \ . 
  \label{RUVa}
\end{align}

For  the link renormalization correction $\Gamma_\Sigma$, 
this Fourier  transform can  be easily done using its original representation (\ref{selfa}),
producing 
   \begin{align} 
     R_\Sigma & (y,x;Pa)= 
\frac {\alpha_s} {2\pi}\, C_F
  \frac1{Pa} \left [ 
  \frac{e^{ -|y-x| Pa }}{(y-x)^2 }  \right.  \nn & \left.  
 -  {\delta (y-x)} \int_{-\infty}^\infty \frac{d\zeta}{(\zeta-x)^2} 
 e^{- |\zeta-x|Pa} \right ] \  . 
    \label{RSig}
 \end{align}
 Note that,  due to the exponential suppression factor,
 the   \mbox{$\zeta$-integral}  accompanying the $\delta (y-x)$ term 
 converges when   \mbox{$\zeta \to \pm \infty$. } As a result, $ R_\Sigma  (y,x;Pa)$ is given 
 by a mathematically 
 well-defined  expression. 
 
 The $1/Pa$ term in Eq. (\ref{RSig}) corresponds to the linear UV divergence.
 Expanding $e^{ -|y-x| Pa }$ in $a$ gives the $1/|y-x|$ term corresponding
 to the logarithmic $\ln (1+z_3^2/a^2)$ UV divergence in Eq. (\ref{selfex}).
 As we have seen, the same  $\ln (1+z_3^2/a^2)$ UV contribution appears 
 in the vertex corrections. 
 Calculating the Fourier transform of $\ln (1+z_3^2/a^2)$ gives 
  \begin{align}  
    R_V  & (y,x;Pa) =  -\frac {\alpha_s} {2\pi}\ C_F 
  \left [  \frac{1}{ |y-x|}e^{-|y-x|Pa} \right. \nn & \left. 
  - {\delta (y-x)}  \int_{-\infty}^\infty \frac{d\zeta}{|y-\zeta|}  e^{- |y-\zeta| Pa} \right ] \   .
 \label{RV}
 \end{align}
  Now we can represent $ R_\Sigma  (y,x;Pa)$  as a sum 
    \begin{align} 
     R_\Sigma & (y,x;Pa)= 
\frac {\alpha_s} {2\pi}\, C_F 
 \left [   {e^{-|y-x|Pa} } \left (  \frac1{Pa (y-x)^2 } +  \frac{1}{ |y-x|}\right )  \right.  \nn & \left.  
 -   {\delta (y-x)} \int_{-\infty}^\infty  d\zeta\, e^{- |\zeta-x|Pa} \left (  \frac1{Pa(\zeta-x)^2} 
 +   \frac{1 }{ |\zeta-x|}
 \right )
 \right ] \nn &+ 
  R_V   (y,x;Pa) 
  \  . 
    \label{RSRV}
 \end{align}
 of the regularized $1/(y-x)^2$ singularity  corresponding to the linear divergence 
 and the vertex kernel $  R_V   (y,x;Pa) $ corresponding to the logarithmic  divergence.

%%%%%%%%%%%%
Here  we want to emphasize that keeping the UV regulator nonzero,
we get mathematically well-defined expressions that produce 
the contribution  $Q_{\rm UV}(y,P)$ satisfying
  \begin{align}
   \int_{-\infty}^\infty   dy \, Q_{\rm UV}(y,P) =  \int_{-\infty}^\infty   dy \,   \int_{-1}^1 dx\,
R_{\rm UV}  (y-x;a) \, f(x) \ =0 , 
  \label{QUVaint}
\end{align} 
 and thus not violating the quark number conservation.
 If one takes $a=0$, the  $\zeta$-integrals in Eqs. 
 (\ref{RSig})  and   (\ref{RV})  diverge when $\zeta \to \pm \infty$,
 and loose mathematical meaning, 
 just like in  expressions given in Ref.  \cite{Xiong:2013bka}.  
 
 One may argue that the limit $a\to 0$ (or equivalent procedure) 
 should be taken to renormalize UV singularities.
 Our answer is that one should simply   consider  
 the reduced ITD introduced in our paper  \cite{Radyushkin:2017cyf} 
 that does not have the UV divergent terms,
 so that no UV-renormalization is needed, and the $a\to 0$ limit poses no problems. 
 Also,  the reduced ITD  does not produce 
the  problematic $\sim 1/\zeta$  terms in the plus-prescription integrals.

As we will show below, the evolution contribution  remaining in the reduced 
ITD, has $\sim 1/y^2$ behavior 
for large $|y|$, and the relevant 
plus-prescription integrals 
are perfectly convergent for large $|\zeta|$.

%%%%%%%%%%%%%%%%%%%%

\subsection{Evolution-related terms}

Let  us  consider now the hard part  given by
 the evolution logarithms 
     \begin{align}
 {\cal  M}_{\rm log}  (\nu, z_3^2) =&   \,    \frac{\alpha_s}{2\pi} \, C_F\,  L_ \Lambda (z_3^2  \Lambda^2 )  
\int_0^1  du \,  B (u) \  {\cal  M}^{\rm soft}  (u \nu, 0)   \  . 
 \label{hardP}
\end{align} 
In this case
 \begin{align}
  Q_{\rm log}  (y, P)  & =
 C_F \, \frac{\alpha_s}{2 \pi}  \, \int_{-1}^1 {d\xi}\, \, f(\xi)  
  \int_{0}^1 du\,  %  
 B(u)  \, K(y-u \xi ;\Lambda^2/P^2) \, 
 \  , 
 \
 \label{Mlog}
\end{align} 
where the kernel $ K(y-u \xi ;\Lambda^2/P^2) $ is given by the  Fourier transform of  $L_ \Lambda (z_3^2  \Lambda^2 ) $.
If we choose the IR regularization of \mbox{Eq. (\ref{Lzm})}  leading to  the 
modified Bessel function $K_0 (mz_3)$,  
we have 
 \begin{align}
 K(y-x,m^2/ P^2) &=\frac{P}{2 \pi}  \int_{-\infty}^\infty dz_3\,
      e^{-i(y-x) Pz_3} \, K_0(m|z_3|) 
 \nonumber \\
  &=  \frac{1}{\sqrt{(y -x)^2+m^2/ P^2}} \  . 
 \ 
 \label{Kyx}
\end{align} 
We may  also write 
 \begin{align}
  Q_{\rm log}  (y, P)  & =
 C_F \, \frac{\alpha_s}{2 \pi}  \, \int_{-1}^1 \, 
 \frac{d\xi} {|\xi|}  R(y/\xi, m^2/ \xi^2P^2)
 \  f(\xi) 
 \  , 
 \
 \label{QR}
\end{align} 
where the kernel $R(\eta, m^2/ P^2) $ \, 
  is given by 
   \begin{align}
 R(\eta;m^2/ P^2)  &= \int_{0}^1 \frac{du }{\sqrt{(\eta  -u )^2+m^2/ P^2}} \left [ \frac{1+u^2}{1-u} \right ]_+
  \  .
 \label{Reta}
\end{align}

%%%%%%%%%%%%%
The kernel $ R(y/\xi, m^2/ \xi^2P^2)$ corresponds to a part 
of the matching factor $Z$ used in the quasi-PDF approach \cite{Ji:2013dva}.
The fact that its mass-dependence comes through the 
combination $m^2/\xi^2 P^2$ was re-discovered in the recent paper
\cite{Izubuchi:2018srq}. 
%%%%%%%%%%%%%%

It is convenient to consider the cases $\xi>0$ and $\xi<0$ separately.
Let us take $\xi>0$ which corresponds to $f(\xi)$ being nonzero for positive $\xi$ only.

\subsubsection{Middle part}

Simply taking $m^2/P^2 = 0 $  results in  a factor $1/|\eta -u|$.
When $0\leq \eta \leq 1$, 
it produces 
 a  non-integrable singularity 
 for $u=\eta$. 
 A more accurate  statement is that, 
with respect to integration over the $0\leq u \leq 1$ interval, we may represent 
     \begin{align}
&  \left. \frac{1 }{\sqrt{(\eta  -u )^2+m^2/ P^2}} \right |_{m^2/P^2 \to 0}= \left ( \frac{1}{|\eta -u|} \right )_+ 
  \nn & +
 \delta (\eta- u)  \ln  \left [ 4\eta (1-\eta) \frac{P^2 }{m^2} \right ] 
  \  , 
 \label{yeta}
\end{align} 
where   the plus-prescription  is defined by 
 \begin{align}
  & \left ( \frac{1}{|\eta-u|} \right )_+ = \frac{1}{|\eta -u|}  - \delta (\eta -u)
 \int_0^1 \frac{d\varv}{|\eta-\varv|} 
  \  . 
\end{align} 
From the $ \delta (\eta -u)$ part we get  the evolution  term 
 \begin{align}
 R_1^{\rm middle} &(\eta; m^2/P^2)  =\ln \left (\frac{4P^2}{m^2} \right )  \,
 \left [ \frac{1+\eta^2}{1-\eta} \theta (0\leq \eta \leq 1 )\right ]_+
 \ ,
 \label{Rmid}
\end{align} 
that is present in the   $0\leq \eta \leq 1$ region only.  
The remaining $\sim \ln \eta (1-\eta )$ term and 
terms  coming  from $(1/|\eta-u|)_+$
 are given   in this region  by 
    \begin{align}
  R_2^{\rm middle } & (\eta) =
 \frac{1+\eta^2}{1-\eta} \log \left [\eta  (1-\eta)  
   \right]
\nn & +\frac{3/2}{ 1-\eta} +4\frac{\log (1-\eta)}{1-\eta} - 1+2 \eta \   . 
\label{central}
   \end{align}

\subsubsection{Outer parts}

For $\eta$ outside the $0\leq \eta \leq 1$ segment,
the $m^2/P^2 \to 0$ limit is finite and  given by 
 \begin{align}
R (\eta ;0)&|_{\eta >1}  = \int_{0}^1 \frac{du }{ \eta - u } \left [ \frac{1+u^2}{1-u} \right ]_+
=- \sum_{n=1}^\infty \frac{\gamma_{n}}{\eta^{n+1}} \ , 
\label{Routp}
\end{align}
   where $\gamma_{n}$ are the anomalous dimensions 
of operators with $n$ derivatives
    \begin{align}
\gamma_{n} =& \int_0^1 du \, \frac{1-u^n}{1-u}  (1+u^2) 
  = 2 \sum_{j=1}^{n+1 }\frac1{j} - \frac32 - \frac{1}{(n+1)(n+2)}  \   .
   \end{align} 
At first sight,  one would expect a $\sim 1/ |\eta|$ behavior for large $|\eta|$.
However, the $1/ |\eta|$  term is accompanied by the integral of $P(u)$ which vanishes  because 
of the plus-prescription. 
This is also the reason why   $\gamma_0$   vanishes causing 
 the series in Eq. (\ref{Routp})  to start at $n=1$. 
In  a closed form,\footnote{Comparing our results with  those of Ref. \cite{Ji:2015jwa},
one should take into account that the evolution-related contributions
are combined there with the \mbox{UV-singular}  terms taken in the limit 
equivalent to our 
$a\to 0$.}
 \begin{align}
R(\eta;0) &|_{\eta >1}  = \frac{1+\eta^2}{\eta-1}\ln
   \left(\frac{\eta-1}{\eta}\right) +
   \frac{3}{2 (\eta-1)}+1 \   . 
   \label{right}
\end{align}
Similarly, for $\eta<0$
 \begin{align}
R(\eta;0) |_{\eta<0} =&\frac{1+\eta^2 }{1-\eta}
\ln 
   \left(\frac{1-\eta}{-\eta}\right) +
   \frac{3}{2 (1-\eta)}-1 \   . 
   \label{left} 
 \end{align}

   One may notice here the    $\pm \frac32 /(1-\eta)$  terms  
   having the evident $\sim 1/|\eta|$ behavior.
   But these terms exactly cancel the $\sim  1/|\eta|$ 
   contributions coming from the remaining terms, thus 
  changing the large-$\eta$ behavior  to  \mbox{$\sim 1/\eta^2$.}
  We have already seen the $\sim - 1/\eta^2$  asymptotic  behavior for $\eta >1$ in
  \mbox{ Eq.(\ref{Routp}).}
 Similarly, for large negative values, one may use the expansion 
 \begin{align}
R(\eta;0) |_{\eta<-1} = \sum_{n=1}^\infty \frac{\gamma_{n}}{\eta^{n+1}} \ . 
\label{Routn}
\end{align}
Thus, for large $|\eta|$ we have the asymptotic behavior
\begin{align}
R(\eta;0) |_{|\eta | \gg 1} =   -\frac43  \frac {{ \rm sgn} (\eta) }{\eta^2} +{\cal O} (1/\eta^3)   \  .
\end{align}

\subsection{Quark  number conservation}

The $\sim 1/y^2$  result  for $R(y/\xi; m^2/P^2)$ may be foreseen if one notices that  calculating  it 
 from the convolution with the AP kernel
(see Eq. (\ref{Reta})), one deals with  the difference 
 \begin{align}
 \frac{1 }{\sqrt{(y -u\xi )^2+m^2/ P^2}} - \frac{1 }{\sqrt{(y -\xi )^2+m^2/ P^2}}
 \   , 
 \label{Ryzmp2}
\end{align} 
which  behaves like $1/y^2$ for large $y$. 
Hence, the   integral of  $ R(y/\xi;m^2/ P^2)$ over $y$ does not have divergences
for large $|y|$. 
Moreover, since the two  terms differ just by a shift in the \mbox{$y$-variable,}
the integral vanishes. As a result, we have
 \begin{align}
 \int_{-\infty}^\infty dy\,  R(y/\xi;m^2/ P^2)   = 0 \ , \
 \label{RKyx}
\end{align} 
which leads to 
 \begin{align}
 \int_{-\infty}^\infty dy\, Q_{\rm log}  (y, P)  = 0 \ , \
 \label{QKyx}
\end{align} 
i.e. the evolution  part does not change the  number of the  valence quarks.

\subsection{$R$-kernel  at finite momenta} 

Results for $R(\eta)$ correspond to the full function
$R(\eta, m^2/P^2)$  taken in the $P^2/m^2  \to \infty $ limit. 
However, since $m$ should be understood as an 
IR cut-off provided by the hadron size, it has a rather large $\sim 0.5$ GeV magnitude.
On the  other hand, the  maximal momenta $P$  reached 
in actual lattice calculations of the quasi-PDFs 
 \cite{Lin:2014zya,Alexandrou:2016jqi}
range from 1.3 to about 2.5 GeV. Thus, it is interesting 
to look at the $P/m$-dependence of  $R(\eta, m^2/P^2)$. 

\begin{figure}[t]
 \centerline{\includegraphics[width=3.3in]{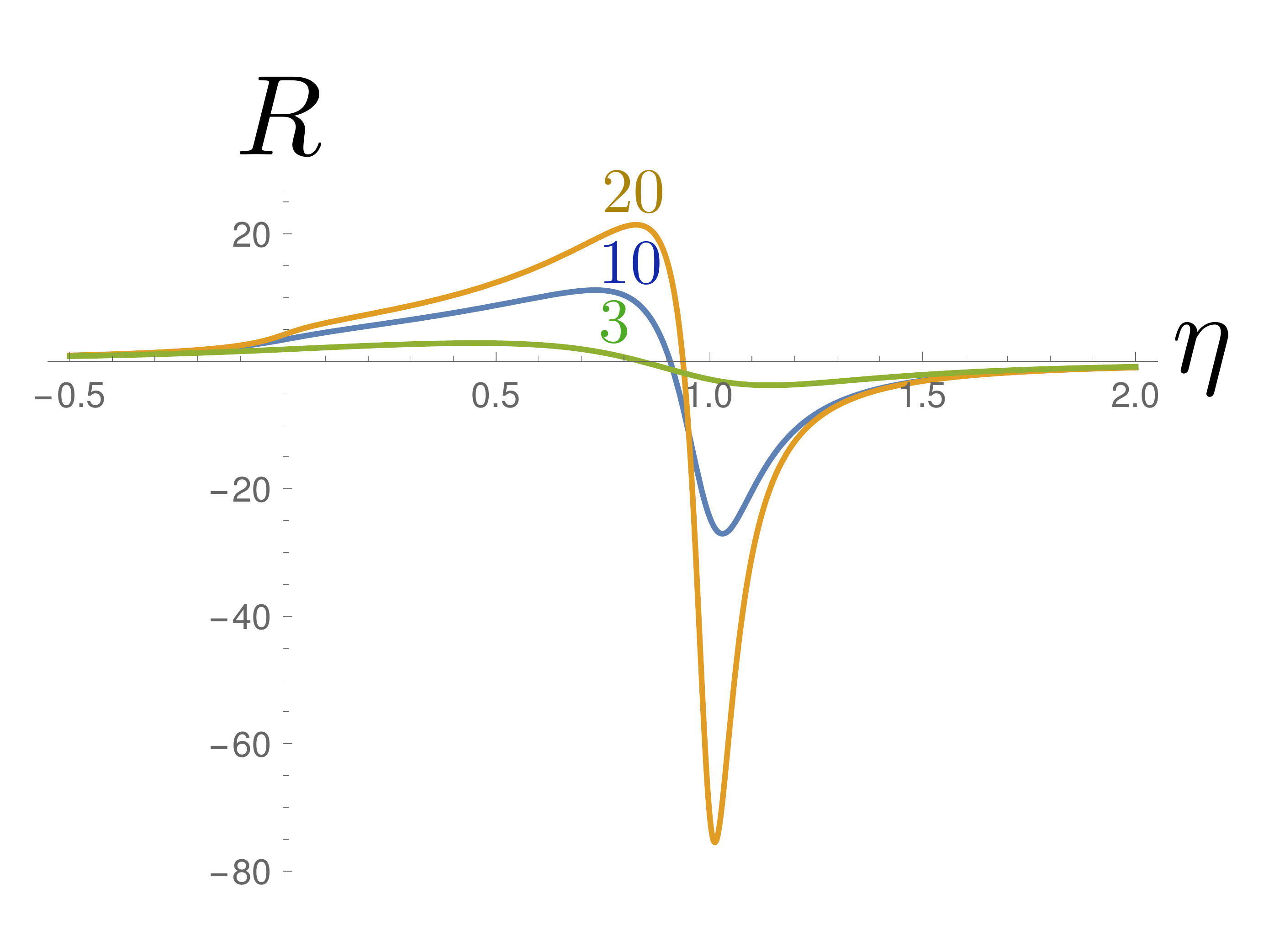}}
    \vspace{-0.2cm}
    \caption{The kernel $R(\eta, m^2/P^2)$   for $P/m=3,  10$ and 20.
    \label{Rm}}
    \end{figure}

 In Fig.  \ref{Rm}, we show the structure of the kernel $R(\eta;m^2/P^2)$ 
  for three  values of $P/m$.  
In the central segment  $0<\eta<1$, it has the evolution part (\ref{Rmid})  proportional 
to $\ln (P^2/m^2)$. This term  is the main reason 
for  the increase of $R$ with $P$  in this region.  A large negative peak 
in the $y\sim 1$ region also increases its magnitude as $\ln (P^2/m^2)$.
It reflects the $\delta (1-\eta)$  plus-prescription term in the AP kernel.

Note that  since  the 
$\ln P^2/m^2$  %%%%%%
%%%%%%  evolution 
part (\ref{Rmid})  has the  plus-prescription  form,
its contribution to the integral  (\ref{RKyx})  is zero.  There are, in addition,  non-logarithmic parts
(\ref{central}), (\ref{right}),  (\ref{left}), and their combined  contribution to the integral  (\ref{RKyx})   should also  vanish. 
This means that in the $P/m \to \infty$ limit the negative peak for $\eta=1$  should contain 
also a non-logarithmic
(in $P^2/m^2$) part that provides plus-prescription for each of these contributions.
For instance, when  $\eta >1$, we would have
 \begin{align}
R(\eta;0) &|_{\eta >1}  \to  R(\eta;0) |_{\eta >1}     -\delta (\eta-1)   \int_1^\infty d \zeta
R(\zeta;0)  
 \   ,
   \label{rightplus}
\end{align}
and similarly for the $0\leq  \eta \leq 1$ and $\eta \leq 0$ parts. %%%%%%%%
The integral over $\zeta$ 
%%%%%%%    here 
for the $\eta\geq 1 $ and $0\leq  \eta \leq 1$ parts %%%%%%%%%
 diverges when $\zeta \to 1$, but this is 
what is expected  from the plus-prescription construction.
For $\zeta \to \infty$, the integral converges,  
hence Eq. (\ref{rightplus}) is a mathematically well-defined expression. 
For the $\eta \leq 0$ part,  the $\zeta$- integral   %%%%%%%%%%
 is just a number. %%%%%%%
 
For nonzero $m/P$, we should have, of course, some smoothened version of $\delta (1-\eta)$
for the $0\leq  \eta \leq 1$ and $\eta \geq 1$ parts.
Looking at the curve corresponding to $P/m=3$
(the value,  realistically corresponding to  the momentum $P\sim 1.5$ GeV),
one can see that ``smoothened''  in this case is a
strong   understatement.
This curve does  not show anything resembling 
 a delta-function near \mbox{$\eta=1$.} 
 The reason for a big difference between 
 the finite-$P$  curves and their $P \to \infty$ limit 
 may be traced to the basic  relation
 (\ref{QyP})  between quasi-PDFs and  pseudo-PDFs:  
  the $y$-shape of quasi-PDFs  $Q(y,P)$
 is  strongly distorted by the large-$z_3$ nonperturbative 
 behavior of the pseudo-PDFs ${\cal P}(x,z_3^2)$. 
 One needs large momenta $P$ (and some deconvolution
 techniques)  just to get rid of this 
 contaminating  large-$z_3$ information.

%%%%%%%%%%%%%%%%%
 In a practical aspect, this means that 
 studying quasi-PDFs at momenta $P$ accessible 
at present-day lattices,  it is a much better approximation
to  ignore  the one-loop corrections  rather than to 
 take  them in the $m/P \to 0$ form.
 %%%%%%%%%%%%%%%%%%%%%
 
In contrast,  if one does not like   to use 
 large $z_3$ values working with  the pseudo-PDFs, one can
simply exclude them  from the analysis.  
Moreover, since the impact parameter distribution
${\cal P}(x,z_\perp^2)$ of the TMD studies 
has the same  functional form as the pseudo-PDF ${\cal P}(x,z_3^2)$, 
one can use  the large-$z_3$  data to get a direct  information
about  the 3-dimensional hadron structure.

\section{ Summary.}

In this  paper, we have studied   the small-$z_3^2$
behavior of the Ioffe-time distribution  ${\cal M} (\nu, z_3^2)$ 
at one-loop level.  
The ITD is the basic function that may be converted,
in a prescribed way,  into 
pseudo-PDFs and quasi-PDFs. 
In its turn, the short-distance structure 
of the ITD is determined by that of the underlying 
bilocal operator ${\cal O}^\alpha (0,z) =\bar \psi (0) \gamma^\alpha E(0,z;A) \psi (z)$.
Since the corrections to ${\cal O}^\alpha (0,z)$ may be calculated 
on  the operator level, it is  an 
even more fundamental object.

In our study, we made an effort to
separate  two sources of the $z_3^2$-dependence 
at small   $z_3^2$. One  is related to the 
UV singularities generated by the gauge  link
$E (0,z_3;A)$  connecting the quark fields 
forming the QCD bilocal operator.
The logarithmic part of these  terms has 
the $\ln (1+z_3^2/a^2)$ structure, where 
$a$ is the UV cut-off parameter analogous to lattice spacing.
 Thus, while being 
singular in the $a\to 0$ limit, the  $\ln (1+z_3^2/a^2)$ factor 
vanishes  for $z_3^2=0$. This is a general property
of the link-related UV-singular terms.  This  property  is  a  very  important one since it 
guarantees  that such corrections do not change the number of valence quarks.

The one-loop UV divergences are  eliminated if 
one considers the reduced ITD ${\mathfrak M} (\nu, z_3^2)$  given 
by the ratio ${\cal M} (\nu, z_3^2)/{\cal M} (0, z_3^2)$. 
Still, ${\mathfrak M} (\nu, z_3^2)$  has a non-trivial short-distance 
behavior. At one loop,  it has the  $\ln z_3^2  \Lambda^2$ structure,
where $\Lambda$ is an IR cut-off parameter. 
These terms generate  perturbative evolution of the parton densities.
 While they are 
 singular in the $z_3^2 \to 0$ limit, the evolution 
 corrections do 
 not change the  number of  valence quarks.
 This is secured by the fact that the $\nu$-dependence 
 of such corrections is governed by factors possessing 
 the plus-prescription property.
The explicit expression that  we give for the $z_3^2$-dependence 
of the reduced ITD at one loop, may (and will)  be used 
in our future work on extraction  of  PDFs from the lattice QCD  simulations
using the pseudo-PDF-based methodology\footnote{These results have been already
used for such a study  in our recent paper \cite{Radyushkin:2018cvn}}. 

We have also demonstrated that our results 
may be used for a   rather   straightforward calculation  
of the 
one-loop corrections to  quasi-PDFs, 
providing new insights concerning   their structure.
In particular, we have demonstrated  that keeping a nonzero %%%%%%%%
UV regulator $a$, one can obtain mathematically well-defined %%%%%%%%
expressions for quasi-PDFs, involving the plus-prescription integrals  %%%%%%%%
that do not diverge at infinity.  %%%%%%%%%

For the UV-finite evolution part, we have demonstrated  that %%%%%%%
they produce quasi-PDFs with $\sim 1/y^2$  behavior that  %%%%%%%%%
also results in convergent plus-prescription integrals. %%%%%%%%%%
We have also observed  that the the mass-dependence of the matching kernel, that  relates  %%%%%%
quasi-PDFs with the ordinary collinear PDFs $f(\xi)$, comes through the $m^2/\xi^2P^2$ combination. %%%%%%%

We have also argued that the IR scale $m$ should be treated as a parameter %%%%%
whose size is of an order of the inverse hadron radius $1/R\sim 0.5$ GeV.  %%%%%%%%
As a result, for presently accessible hadron momenta $P$,  %%%%%%%%%%
the one-loop corrections for quasi-PDFs are  much smaller than in the formal $m/P \to 0$ limit. %%%%%%%%%%%

As emphasized above,  the corrections to the bilocal 
operator ${\cal O}^\alpha (0,z)$ may be calculated 
 without specifying a matrix element
in which it is embedded.  In particular, changing the  $\langle p| ...|p\rangle$
brackets into $\langle 0| ...|p\rangle$, one may use the results of the present paper
to get  one-loop corrections to pseudo- and quasidistribution amplitudes.
Similarly, taking the $\langle p_1| ...|p_2\rangle$  matrix elements, 
one can get one-loop  results for generalized parton pseudodistributions
(``pseudo-GPDs''). These are natural directions for  future studies.

{\bf Acknowledgements.} I thank I. Balitsky, V.M. Braun, L. Jin,   \mbox{J.-W. Qiu} and Y. Zhao   for discussions and
K. Orginos    for  his 
 interest in  this work.   
This work is supported by Jefferson Science Associates,
 LLC under  U.S. DOE Contract \#DE-AC05-06OR23177
 and by U.S. DOE Grant \#DE-FG02-97ER41028.

\end{document}